# Plasmonically induced transparent magnetic resonance in a metallic metamaterial composed of asymmetric double bars

Zheng-Gao Dong,[1,a)] Hui Liu,[2] Tao Li,[2] Shu-Ming Wang,[2] Shi-Ning Zhu,[2] and X. Zhang[3]

[1]*Physics Department, Southeast University, Nanjing 211189, China*

[2]*National Laboratory of Solid State Microstructures, Nanjing University, Nanjing 210093, China*

[3]*5130 Etcheverry Hall, Nanoscale Science and Engineering Center, University of California, Berkeley, California 94720-1740, USA*

We demonstrate that the trapped magnetic resonance mode can be induced in an asymmetric double-bar structure for electromagnetic waves normally incident onto the double-bar plane, which mode otherwise cannot be excited if the double bars are equal in length. By adjusting the structural geometry, the trapped magnetic resonance becomes transparent with little resonance absorption when it happens in the dipolar resonance regime, a phenomenon so-called plasmonic analogue of electromagnetically induced transparency. Accordingly, the additional magnetic resonance by introducing asymmetry is explained as a result of nonlinear plasmon coupling.

For conventional metamaterials, the spatial distances between discrete elements were generally out of the touch of neighboring near fields localized around individual elements, and consequently the plasmonic interactions of nearby metal structures could be neglected. Though this treatment was feasible to study the average effect in terms of the effective medium approximation (such as the left-handed metamaterials), it is not applicable to interparticle-coupled metamaterials, for which great interest has been provoked recently because the plasmon coupling between adjacent metallic elements can induce many attractive electromagnetic properties.[1-3] For example, the plasmon hybridization in neighboring elements can split the resonant spectrum and obtain a great optical activity,[4] while the plasmonic analogue of electromagnetically induced transparency (EIT) in metamaterials, usually composed of dual resonators,[5-11] is a result of plasmon coupling between a radiative eigenmode (e.g., dipolar resonance) in one resonator and a subradiant eigenmode (e.g., quadrupolar resonance) in the adjacent resonator in a manner of destructive interference.

Generally, the intriguing properties in coupled metamaterials are resulted from various plasmon coupling configurations with either structural symmetry or asymmetry. A straight configuration is to squeeze the element interval and thus the near-field coupling can be significantly enhanced between metal elements, including those symmetric elements equivalent in both metal shape and element arrangement.[12] In contrast, spatially and/or structurally asymmetric configurations usually take astonishing roles in modifying the electromagentic responses in coupled metamaterials. Spatially asymmetric coupling, i.e., rotating or translating the neighboring homomorphic elements with respect to one another can lead to optical activity or additional dark-mode excitation, respectively.[4,13] On the other hand, structurally asymmetric coupling happens by deliberately breaking the element symmetry in shape as well as in size, such as the concentric double rings,[14,15] ring-disk composite,[16] asymmetric split-ring pairs,[17,18] and mismatched nanoparticle pairs.[19] A common coupling characteristic for such size or shape asymmetry in adjacent elements is the Fano-type resonance with a transmission dip closely accompanied with a transmission peak.[14-18,20] In this work, we numerically demonstrate that a trapped magnetic/quadrupolar resonance with Fano-type profile is excitable by introducing asymmetry in a double-bar structure. A plasmonically induced transparency can be obtained when this trapped magnetic resonance coincides with a dipolar resonance. Based on the interpretation of the plasmonic analogue of EIT phenomenon, it is considered that this asymmetry induced quadrupolar resonance is a result of nonlinear plasmon coupling (i.e., frequency-conversion plasmon coupling).

Figure 1 shows a schematic illustration of the metallic metamaterial and the polarization of incident light. The bar geometry with length $l = 200\,\text{nm}$, width $w = 30\,\text{nm}$, thinckness $t = 30\,\text{nm}$, and the gap between double bars is $g = 30\,\text{nm}$. The translating parameters for the double bars are: $p_x = 250\,\text{nm}$, $p_z = 100\,\text{nm}$, and $p_y = 600\,\text{nm}$. The substrate is quartz with index of refraction $n_s = 1.55$, and the metallic bars are immerged in a host medium with index of refraction $n_h = 1.94$. Under the polarization situation of incident waves with electric field along the bar length, dipolar oscillations are excitable bright mode, whereas the quadrupolar eigenmode with antiparallel induced currents inherent to the double parallel bars is subradiant/dark because the double bars with a same length are electrically equivalent. In our simulations,[21,22] perfect electric and magnetic boundaries are used in compliance with the incident polarization configuration, and the metal is defined by using Drude dispersion with $\omega_p = 1.37 \times 10^{16} \text{s}^{-1}$ and $\gamma = 1.2 \times 10^{14} \text{s}^{-1}$, where $\gamma$ is three times of the theoretical bulk value in order to account for nanofabrication factors.[7,8]

Figures 2(a) and 2(b) shows the calculated transmission spectra for the double-bar structures with equal bar lengths $l = 200 \text{ nm}$ and $120 \text{ nm}$, respectively. It is found that the dipolar resonances occur near $f_1 = 162 \text{ THz}$ and $f_2 = 248 \text{ THz}$, respectively. As is known, the asymmetric mode (i.e., the quadruapolar resonant mode with antiparallel induced current at the surface of the double-bar structure) can not be excited magnetically since the magnetic field component is parallel to the double-bar plane, also it can not be excited electrically because the symmetric double bars (i.e., bars in the same length) are equivalent with respect to the polarized electric field. However, by breaking the length symmetry of the double bars, e.g., keeping the length of one of the double bars to $l = 200 \text{ nm}$ while shortening the length of another bar to be $l' = 120 \text{ nm}$, a third resonant mode, in addition to the two dipolar resonances as presented in Figs. 2(a) and 2(b), can be induced around $f_3 = 202 \text{ THz}$ with a Fano-type transmission spectrum, as shown in Fig. 2(c). This additional mode confines the strong magnetic field in the gap of the asymmetric double bars, and thus is a nonradiative/trapped magnetic resonance with antiparallel induced currents on the double bar surfaces. To confirm these dipolar and quadrupolar modes, the magnetic field distributions are presented in Figs. 2(d)-2(h), corresponding to the five resonances in Figs. 2(a)-2(c).

Before explaining the physical origin of the additional magnetic resonance for the asymmetric double-rod structure, we show that a plasmon version of the EIT phenomenon can be obtained in this asymmetric double-bar metamaterial, if the structural parameters are adjusted to make the magnetic resonance locate within the frequency regime of the dipolar resonance. As shown in Fig. 3, when the dipolar and quadrupolar resonances coincide, the latter one will become transparent. In different to the EIT phenomenon in an atomic system, the intrinsic metal loss in the metamaterial can not be eliminated at the optical spectrum. This transparency window, as well as the corresponding narrow dip at the middle portion of the dipolar absorption profile, is a result of the destructive interference between the two excitation pathways, namely, direct excitation of the radiative dipolar plasmon oscillation and indirect excitation of the nonradiative quadrupolar plasmon oscillation.[5,7]

According to the interpretation for a plasmonic analogue of EIT phenomenon,[4-8] the incident waves excite the dipolar plasmon state, and then the excited dipolar oscillation plasmonically couples to the quadrupolar oscillation. Transparency window can be formed if there is a destructive interference between the direct excitation pathway of dipolar plasmon oscillation and the indirect excitation pathway of quadrupolar plasmon oscillation. During this process, radiative plasmon state transfers the exciting electromagnetic resonance to the subradiant plasmon state through the aid of near-field plasmon coupling. In this sense, for the physical origin of the additional magnetic resonance in the asymmetric double-bar structure [Fig. 2(c)], we can consider that there is a nonlinear plasmon coupling where the dipolar modes at $f_1$ and $f_2$ are coupled into a quadrupolar mode at $f_3$. Namely, a frequency-conversion coupling exists between these plasmon eigenmodes. It is the underlying plasmon coupling that makes the otherwise dark quadrupolar mode excitable. Although this nonlinear plasmon coupling is striking with respect to the same-frequency plasmon coupling in EIT-like metamaterials,[5-11] the optical nonlinearity in EIT mechanism is inherent in atomic system where a control light is different to a probe light in frequency.[23,24] Here, the nonlinear plasmon coupling acts as the control factor to decide whether the additional resonance is excitable or not, and thus determines the absorption or transparency of the probe/incident light. Note that the EIT-like peak can not be obtained in the symmetric double-bar structure because no coupling excitation would happen between the dipolar and quadrupolar mode for such case [Figs. 2(a) and (b)], unless a length asymmetry is introduced.

Another characteristic of EIT-like resonance in metamaterials is the large group index that is useful for slowing down the electromagnetic propagation in nanoplasmonic devices. To characterize this property for the EIT-like transmission peak in the asymmetric double-bar metamterial as presented in Fig. 3, the group index dispersion is calculated. It is found in Fig. 4 that the maximum group index can reach a value as large as 27 at the transparency window. In contrast, for the dipolar resonance the retrieved group index in *negative* value should have no significant meaning for slowing light, since the transmission is forbidden in this resonance regime.

In summary, interaction between the radiative dipolar mode and the subradiant quadrupolar mode is numerically investigated in a metallic double-bar metamaterial. The radiative dipolar plasmon state can not only absorb the polarized light, but also it can evoke the subradiant quadrupolar plasmon mode through the near-field plasmon coupling. In addition to the linear plasmon coupling between same-frequency modes which leads to the plasmonic analogue of EIT in the case of destructive interference, we demonstrate in this work that nonlinear plasmon coupling can lead to an additional magnetic resonance excitable at a different frequency. In contrast to a radiative mode directly excited by the incident waves, this coupling-induced mode is narrow with Fano-type profile and can only be excited by breaking the structural symmetry.

**Acknowledgments**

This work was supported by the National Natural Science Foundation of China (No.10874081, No.10904012, and No. 60990320), and the Research Fund for the Doctoral Program of Higher Education of China (No. 20090092120031).

**References**

[1] H. Liu, Y. M. Liu, T. Li, S. M. Wang, S. N. Zhu, and X. Zhang, "Coupled magnetic plasmonsin metamaterials," Phys. Status Solidi B **246**(7), 1397-1406 (2009).

[2] M. Decker, S. Linden, and M. Wegener, "Coupling effects in low-symmetry planar split-ring resonator arrays," Opt. Lett. **34**(10), 1579-1581 (2009).

[3] T. Li, H. Liu, F. M. Wang, Z. G. Dong, S. N. Zhu, and X. Zhang, "Coupling effect of magnetic polariton in perforated metal/dielectric layered metamaterials and its influence on negative refraction transmission," Opt. Express **14**(23), 11155-11163 (2006).

[4] H. Liu, D. A. Genov, D. M. Wu, Y. M. Liu, Z. W. Liu, C. Sun, S. N. Zhu, and X. Zhang, "Magnetic plasmon hybridization and optical activity at optical frequencies," Phys. Rev. B **76**(7), 073101 (2007).

[5] S. Zhang, D. A. Genov, Y. Wang, M. Liu, and X. Zhang, "Plasmon-induced transparency in metamaterials," Phys. Rev. Lett. **101**(4), 047401 (2008).

[6] N. Papasimakis, V. A. Fedotov, N. I. Zheludev and S. L. Prosvirnin, "Metamaterial analogy of electromagnetically induced transparency," Phys. Rev. Lett. **101**(25), 253903 (2008).

[7] N. Liu, L. Langguth, T. Weiss, J. Kastel, M. Fleischhauer, T. Pfau, and H. Giessen, "Plasmonic electromagnetically induced transparency at the Drude damping limit," Nature Mater. **8**(9), 758-762 (2009).

[8] N. Liu, T. Weiss, M. Mesch, L. Langguth, U. Eigenthaler, M. Hirscher, C. Sonnichsen, and H. Giessen, "Planar metamaterial analogue of electromagnetically

induced transparency for plasmonic sensing," Nano Lett. **10**(4), 1103-1107 (2010).

[9]V. Yannopapas, E. Paspalakis, and N. V. Vitanov, "Electromagnetically induced transparency and slow light in an array of metallic nanoparticles," Phys. Rev. B **80**(3), 035104 (2009).

[10]P. Tassin, L. Zhang, T. Koschny, E. N. Economou, and C. M. Soukoulis, "Low-loss metamaterials based on classical electromagnetically induced transparency," Phys. Rev. Lett. **102**(5), 053901 (2009).

[11]P. Tassin, L. Zhang, T. Koschny, E. N. Economou, and C. M. Soukoulis, "Planar designs for electromagnetically induced transparency in metamaterials," Opt. Express **17**(7), 5595-5605 (2009).

[12]I. Sersic, M. Frimmer, E. Verhagen, and A. F. Koenderink, "Electric and magnetic dipole coupling in near-infrared split-ring metamaterial arrays," Phys. Rev. Lett. **103**(21), 213902 (2009).

[13]A. Christ, O. J. F. Martin, Y. Ekinci, N. A. Gippius, and S. G. Tikhodeev, "Symmetry breaking in a plasmonic metamaterial at optical wavelength," Nano Lett. **8**(8), 2171-2175 (2008).

[14]Z. G. Dong, M. X. Xu, S. Y. Lei, H. Liu, T. Li, F. M. Wang, and S. N. Zhu, "Negative refraction with magnetic resonance in a metallic double-ring metamaterial," Appl. Phys. Lett. **92**(6), 064101 (2008).

[15]N. Papasimakis, Y. H. Fu, V. A. Fedotov, S. L. Prosvirnin, D. P. Tsai, and N. I. Zheludev, "Metamaterial with polarization and direction insensitive resonant transmission response mimicking electromagnetically induced transparency," Appl.

Phys. Lett. **94**(21), 211902 (2009).

[16]F. Hao, Y. Sonnefraud, P. V. Dorpe, S. A. Maier, N. J. Halas, and P. Nordlander, "Symmetry breaking in plasmonic nanocavities: subradiant LSPR sensing and a tunable Fano resonance," Nano Lett. **8**(11), 3983-3988 (2008).

[17]C.-Y. Chen, I.-W. Un, N.-H. Tai, and T.-J. Yen, "Asymmetric coupling between subradiant and superradiant plasmonic resonances and its enhanced sensing performance," Opt. Express **17**(17), 15372-15380 (2009).

[18]R. Singh, C. Rockstuhl, F. Lederer, and W. Zhang, "Coupling between a dark and a bright eigenmode in a terahertz metamaterial," Phys. Rev. B **79**(8), 085111 (2009).

[19]L. V. Brown, H. Sobhani, J. B. Lassiter, P. Nordlander, and N. J. Halas, "Heterodimers: plasmonic properties of mismatched nanoparticle pairs," ACS Nano **4**(2), 819-832 (2010).

[20]V. A. Fedotov, M. Rose, S. L. Prosvirnin, N. Papasimakis, and N. I. Zheludev, "Sharp trapped-mode resonances in planar metamaterials with a broken structural symmetry," Phys. Rev. Lett. **99**(14), 147401 (2007).

[21]Z.-G. Dong, H. Liu, T. Li, Z.-H. Zhu, S.-M. Wang, J.-X. Cao, S.-N. Zhu, and X. Zhang, "Optical loss compensation in a bulk left-handed metamaterial by the gain in quantum dots," Appl. Phys. Lett. **96**(4), 044104 (2010).

[22]F. M. Wang, H. Liu, T. Li, Z. G. Dong, S. N. Zhu, and X. Zhang, "Metamaterial of rod pairs standing on gold plate and its negative refraction property in the far-infrared frequency regime," Phys. Rev. E **75**(1), 016604 (2007).

[23]M. Fleischhauer, A. Imamoglu, and J. P. Marangos, "Electromagnetically induced

transparency: Optics in coherent media," Rev. Mod. Phys. **77**(2), 633-673 (2005).

[24]C. L. G. Alzar, M. A. G. Martinez, and P. Nussenzveig, "Classical analog of electromagnetically induced transparency," Am. J. Phys. **70**(1), 37-41 (2002).

**Figure captions:**

Fig. 1. (Color online) Structural schematic of the metallic double-bar metamaterial. Only single double-bar layer in the propagation direction is considered in the simulations.

Fig. 2. (Color online) Transmission spectra of the double-bar metamaterial with (a) symmetric bar lengths $l = l' = 200$ nm, (b) symmetric bar lengths $l = l' = 120$ nm, and (c) asymmetric bar lengths $l = 200$ nm and $l' = 120$ nm. The magnetic field distributions for the five resonances are shown in the right panel (d-h).

Fig. 3. (Color online) EIT-like transmittance and absorbance in the asymmetric double-bar metamaterial with some adjusted structural scales as follows: $l = 157$ nm, $l' = 117$ nm, $g = 40$ nm, $P_x = 240$ nm, and $P_y = 590$ nm. The orange vertical shadow indicates the position of the transparency window and absorption dip and the inset presents the localized magnetic field distribution of the transparency window.

Fig. 4. Group index of the EIT-like asymmetric double-bar metamaterial with structural parameters as presented in Fig. 3.

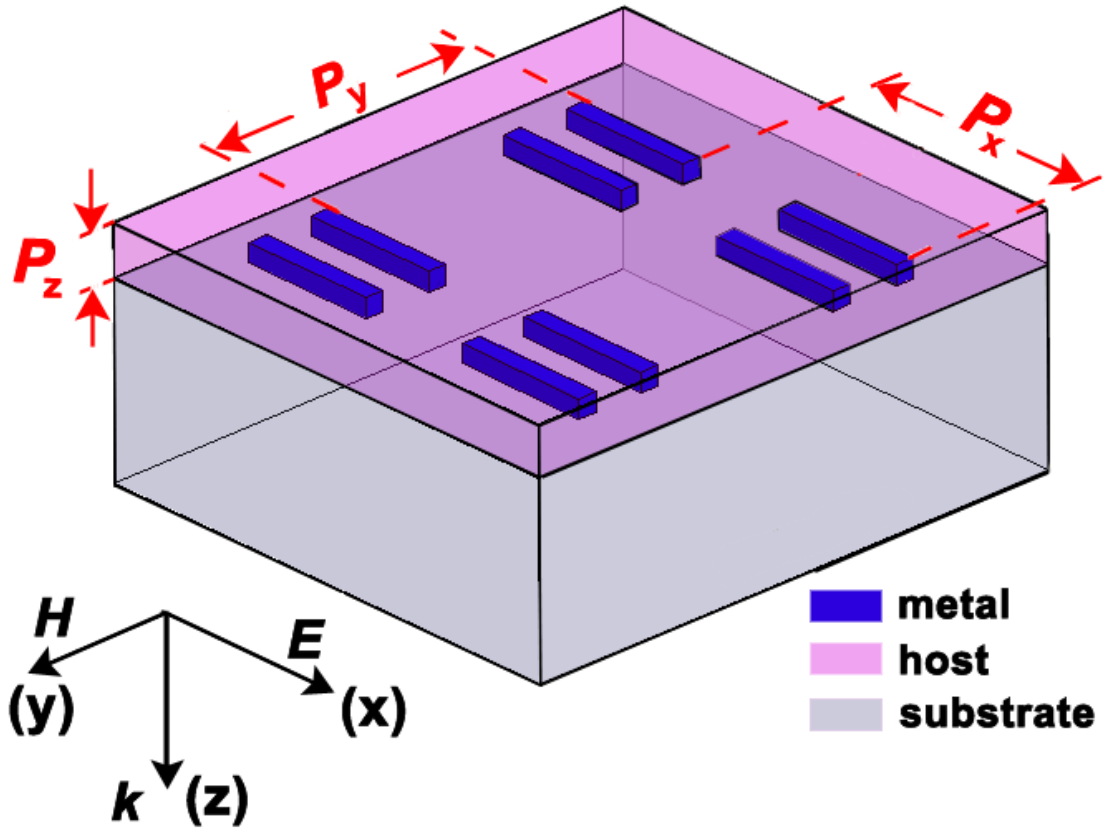

Fig. 1 Dong et al.

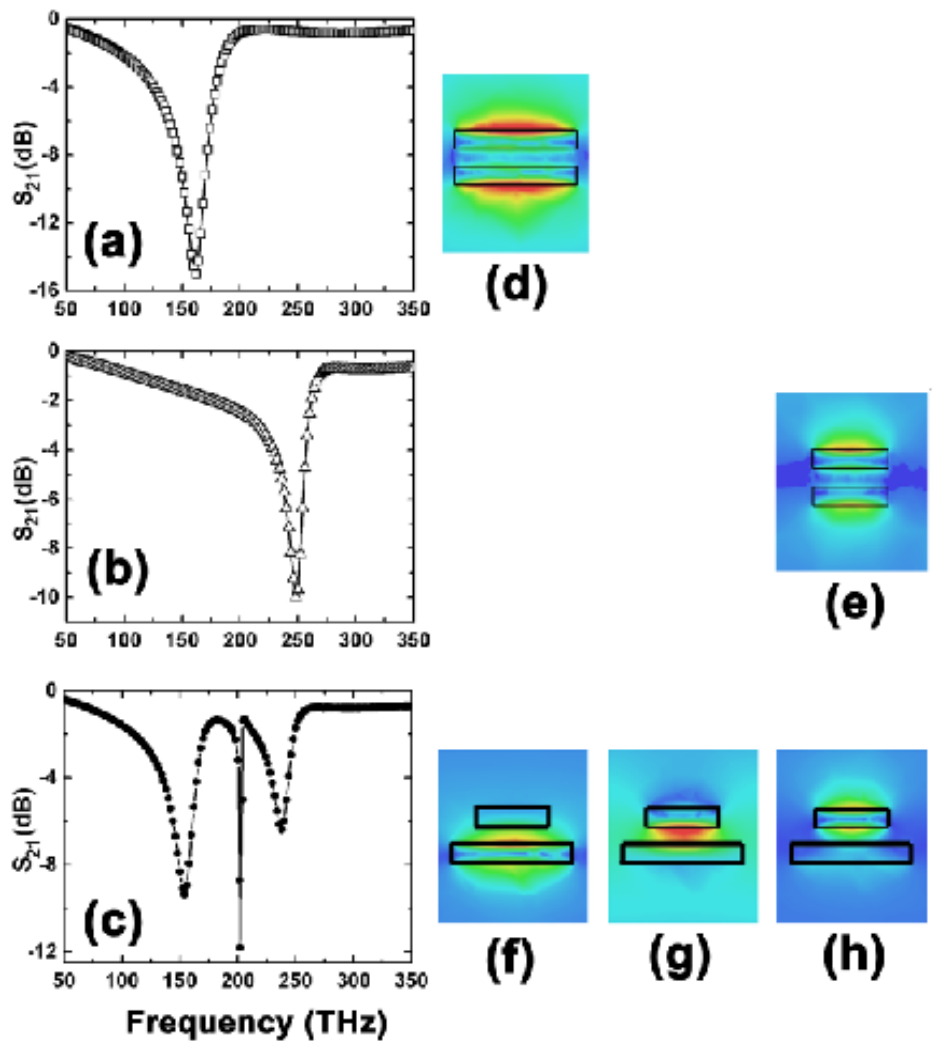

Fig. 2 Dong et al.

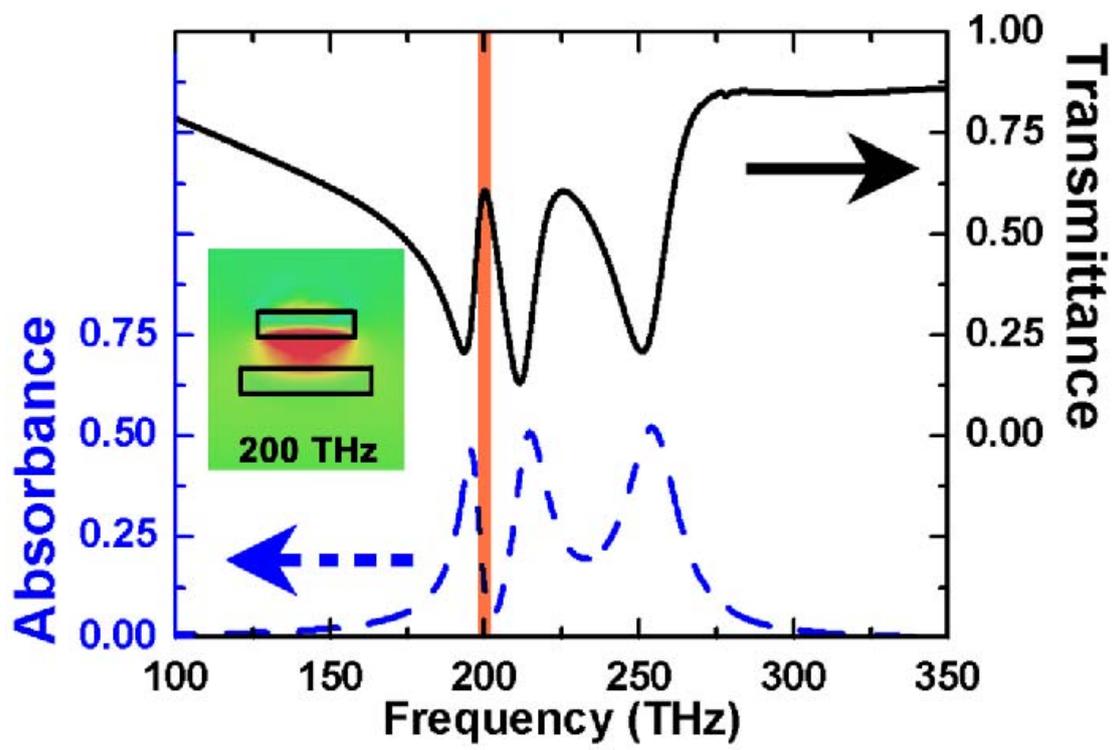

Fig. 3 Dong et al.

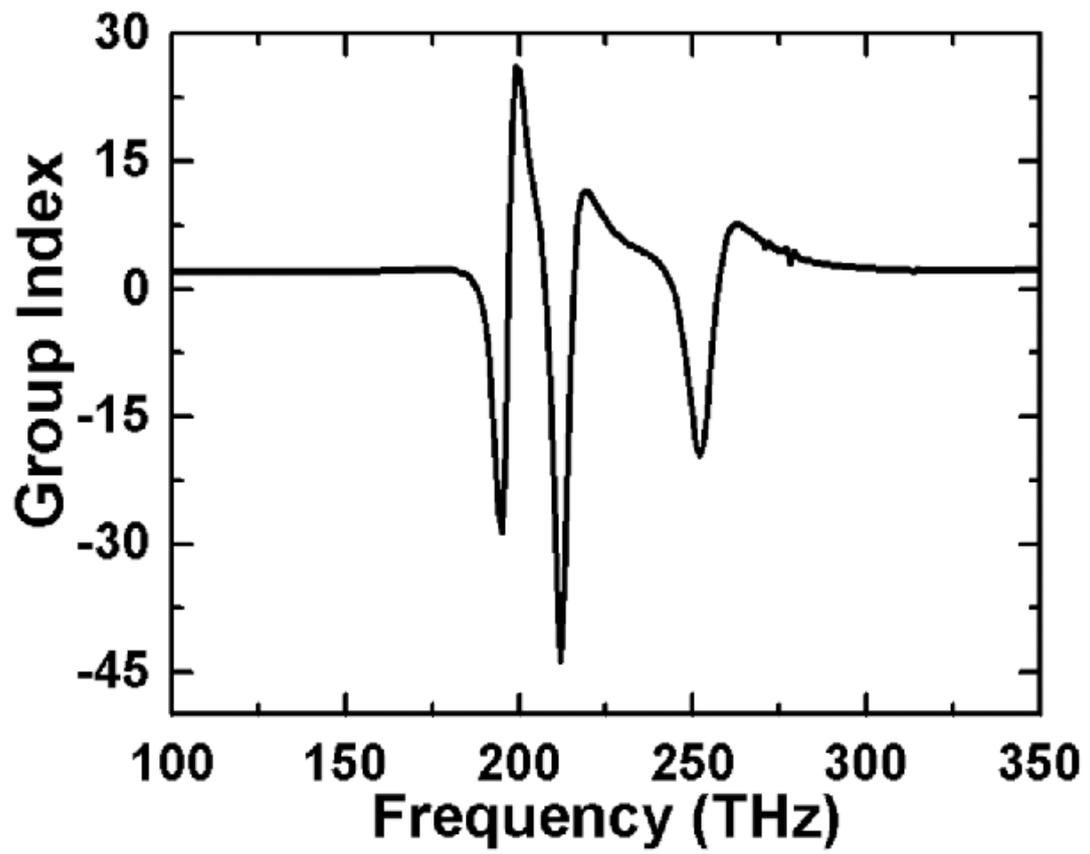

Fig. 4 Dong et al.